\documentclass[a4paper]{article}
\usepackage{amsmath}
\usepackage{amssymb}
\usepackage{graphicx}
\usepackage{authblk}

\begin{document} 

\title{Spectral features in the cosmic ray fluxes}

\author{Paolo Lipari}
\affil{\footnotesize INFN sezione Roma ``Sapienza''} 
%\affil{{\textit paolo.lipari@roma1.infn.it}}

\maketitle

\begin{abstract}
  The cosmic ray  energy distributions  contain  spectral features, that is narrow
  energy regions where the slope of the spectrum changes rapidly.  The identification and study of
  these features  is of great importance to understand the  astrophysical  mechanisms
  of acceleration and propagation that form the spectra.
  In first approximation a  spectral  feature in often  described
  as a  discontinuous  change in slope,   however  very  valuable information
  is also  contained in its width,  that is the length of the  interval in 
  logarithm of energy   where the change in  spectral  index  develops.
  In this  work we  discuss the best way to define and
  parametrize  the  width  a  spectral feature, and  for illustration
  discuss  some  of the most prominent known structures.
\end{abstract}

\section{Introduction}
\label{sec:introduction} 
The spectra of cosmic rays (CR) extend to a very broad energy range with
a smooth shape that, for energy $E \gtrsim 30$~GeV,
is usually described as an ensemble of adjacent energy intervals,
where the energy distribution  is
a simple power law ($\phi(E) \simeq K \; E^{-\alpha}$), 
separated by ``spectral features'', that is narrow regions where the slope
(or spectral index) of the flux undergoes a rapid change.
The  features can be  softenings 
or hardenings of the spectrum, and appear as ``knee--like'' or ``ankle--like''
in the usual log--log graphic representation of the spectrum.
Prominent and well known  examples of features  in the all particle spectrum
are in  fact the ``Knee at $E \simeq 3$~PeV, and the ``Ankle'' at $E \simeq 4$~EeV.

The simple description outlined above is an  approximation,
because it is likely that the CR spectra are
not,  even in  a limited range of energy,  exactly of power law form, 
and  the spectral indices are always slowly  evolving with energy;
however the identification and study of discrete  spectral  features
can be considered as a natural and useful  task.

It is obviously very desirable, and in fact ultimately necessary,
to describe the CR spectral features
in the framework of astrophysically motivated models, 
and in terms of  parameters that have a real physical meaning,
and in the literature there are several alternative models
to interpret the  observations.  On the other hand, it is 
useful to have a purely phenomenological description
of  the shape  of the spectral features, as an intermediate  step
that can be used as a guide in the construction of astrophysical models.

In  first order approximation, a spectral feature 
can be  described as infinitely narrow, with the spectral index
that changes discontinuosly. In this limit  a feature it  is
completely described by four parameters:
$E_b$  the break energy, that gives its position, 
$\alpha_1$ and $\alpha_2$ the spectral slopes
before and after the break, and the absolute normalization of the the flux.

\begin{figure}[bt]
\begin{center}
\includegraphics[width=14.0cm]{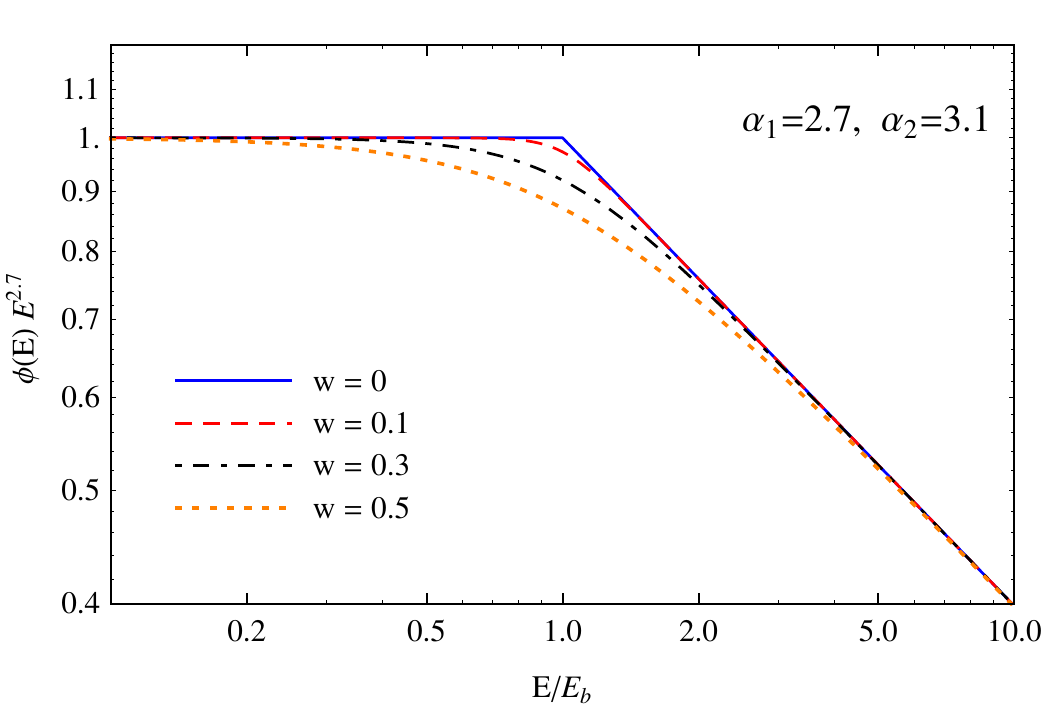}
\end{center}
\caption {\footnotesize
  Example of a (softening) spectral  feature   described by
  the parametrization of Eq.~(\ref{eq:break-parametrization}).
  The spectral indices before and after the break are
  $\alpha_1 = 2.7$ and $\alpha_2 = 3.1$.
  The different curves are for $w = 0$, 0.1, 0.3 and 0.5.
\label{fig:break0}}
\end{figure}

\begin{figure}[bt]
\begin{center}
\includegraphics[width=14.0cm]{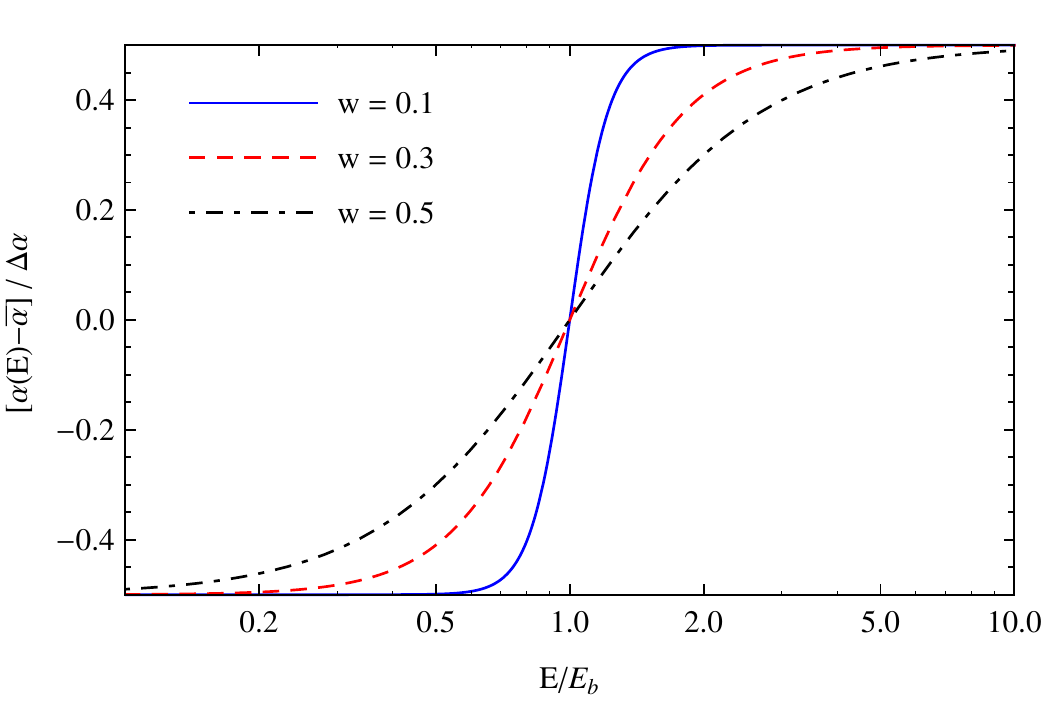}
\end{center}
\caption {\footnotesize
 Energy dependence of the spectral index  [see  Eq.~(\ref{eq:index-parametrization})].
 The three curves correspond to three values of the width parameter
 ($w=0.1$, 0.3 and 0.5).
\label{fig:break}}
\end{figure}

It is obvious that the hypothesis of a discontinuous
change in spectral slope is unphysical, and
this suggests that a  phenomenological description of
a spectral feature should include at least one additional parameter.
A simple and convenient parametrization of the spectral shape
of the CR all particle spectrum   in the region of the  
Knee has been introduced by Ter--Antonyan and  Haroyan \cite{TerAntonyan:2000hh}
and later used by Schatz \cite{Schatz:2001nf}. This parametrization can be
applied to the description of both softening and hardening spectral features
and   (with $E_0$ is an arbitrary  reference energy)  has the form:
\begin{equation}
 \phi(E) = K_0 \;
 \left ( \frac{E}{E_0} \right )^{-\alpha_1} \;
 \left [1 + \left (\frac{E}{E_b} \right )^{\frac{1}{w}} \right ]^{-(\alpha_2 - \alpha_1) \, w}
\label{eq:break-parametrization} 
\end{equation}
that contains one additional parameter, the width  $w >0$
(note that the  authors of  \cite{TerAntonyan:2000hh,Schatz:2001nf}  use the 
parameter $\varepsilon=1/w$).
Some examples of the spectral shapes  of this  parametrization are  shown in Fig.~\ref{fig:break0}.
For a  more precise understanding of the ``geometrical meaning'' of
$w$ it  is useful to consider the energy dependence of the spectral index
of a flux described by  Eq.~(\ref{eq:break-parametrization}):
\begin{equation}
 \alpha(E) \equiv - \frac{d\ln \phi}{d\ln E}
 = \overline{\alpha} + \frac{\Delta \alpha}{2} \;
 \tanh \left [ \frac{\ln (E/E_b)}{2 w} \right ]~.
\label{eq:index-parametrization}
\end{equation}
In this equation
$\overline{\alpha} = (\alpha_2 + \alpha_1)/2$ is the average of the two spectral
indices before and after the break, and
$\Delta \alpha = (\alpha_2 - \alpha_1)$ is the total change in spectral index
across the break (some numerical examples are shown in Fig.~\ref{fig:break}).
It is straightforward to see that  $w$  gives the width
of the energy range where the step in spectral index develops.

The limit $w \to 0$ of  Eq.~(\ref{eq:index-parametrization}) is:
\begin{equation}
 \lim_{w\to 0} \alpha (E) =
 \begin{cases}
 \overline{\alpha} - \frac{\Delta \alpha}{2} = \alpha_1 & \text {for $E < E_b$}
 \\[0.25 cm]
 \overline{\alpha} + \frac{\Delta \alpha}{2} = \alpha_2 & \text {for $E > E_b$} 
 \end{cases}
\end{equation}
and corresponds to a discontinuous jump of the spectral index.
More in general, one has that the asymptotic values (for $E \to 0$ and $E\to \infty$)
of the spectral  index  are $\alpha_1$ and $\alpha_2$, and 
at the break energy $E_b$ the spectral index takes the average value:
$\alpha(E_b) = \overline{\alpha}$.
The jump $\Delta \alpha$ develops symmetrically in $\log E$, and 
the energies $E_{f_\pm}$ where the spectral index takes the values:
\begin{equation}
 \alpha(E_{f_\pm}) = \overline{\alpha} \pm 
 \frac{\Delta \alpha}{2} \; f
\end{equation}
(with $0 \le f <  1$) are given by:
\begin{equation}
\log E_{f_\pm} = \log E_b \pm w \; \log \left [ \frac{1+ f}{1-f} \right ] ~,
\end{equation}
so that the two values $\log E_{f_\pm}$ are placed symmetrically
with respect to $\log E_b$.
The total range  of $\log E$ (centered on $\log E_b$)
where the spectral index varies by $\Delta \alpha/2$ is then: 
\begin{equation}
 (\Delta \log_{10} E )_{\Delta \alpha/2}
% = \frac{2 \, \ln3}{\ln 10} \; w \simeq 0.954 \; w ~.
 = (\log_{10} 9)  \; w \simeq 0.954 \; w ~.
\label{eq:weff}
\end{equation}
This allows to attribute a simple and easy to remember physical meaning
to $w$. The value $w \simeq 1$ corresponds to 
a spectral feature that develops in approximately a decade of energy,
and a feature of width $w \simeq 0.1$
has an energy extension that is approximately
a factor $\approx 10^{0.1} \simeq 1.25$.

Recently the AMS02 collaboration has presented fits to the rigidity spectra of the
proton an helium spectra using the parametrization
(expressed here as a function of energy):
\begin{equation}
 \phi(E) = K \;
 \left ( \frac{E}{E_0} \right )^{-\alpha_1} \;
 \left [1 + \left (\frac{E}{E_b} \right )^{-(\alpha_2 - \alpha_1)/s} \right ]^{s} ~.
\label{eq:break-par-ams} 
\end{equation}
Eqs.~(\ref{eq:break-parametrization}) and~(\ref{eq:break-par-ams}) are
in fact different parametrizations of the same ensemble of curves.
The parameter $s$  used in  Eq.~(\ref{eq:break-par-ams})  is 
related to  the width $w$ of Eq.~(\ref{eq:break-parametrization}) by:
\begin{equation}
s = \frac{w}{|\Delta \alpha|} ~,
\label{eq:s-w}
\end{equation}
and therefore Eqs.~(\ref{eq:break-parametrization}) and~(\ref{eq:break-par-ams}) are  equivalent.
However, we find that the use of the width parameter $w$ is
is preferable because of its more transparent and intuitive physical meaning.
In addition, when performing fits to data, 
the quantities in the pair \{$s$, $\Delta \alpha$\} are in general
much more strongly correlated than the quantities in the pair \{$w$, $\Delta \alpha$\}.

As discussed above, the spectral  index
of a flux described by  Eq.~(\ref{eq:break-parametrization}) or~(\ref{eq:break-par-ams})
is symmetric in $\log E$.
It is potentially interesting to have a more flexible
functional form to describe a spectral  feature
that allows for the possibility that the spectral index
changes more rapidly before of after the break energy.
A simple generalization of Eq.~(\ref{eq:break-parametrization})   that  depends
on one more parameter, can be obtained, keeping  for $E_b$ the 
same definition, that is the energy  where the spectral index
takes the average value:  $\alpha (E_b) = (\alpha_1+\alpha_2)/2$,
and  introducing two different  widths   to the left and right of
the break energy. This   results in the form: 
\begin{equation}
 \phi(E) = \begin{cases} 
 K_0\;
 \left ( \frac{E}{E_0} \right )^{-\alpha_1} 
 \left [1 + \left (\frac{E}{E_b} \right )^{\frac{1}{w_L}} \right ]^{ -\Delta \alpha \, w_L} &
 \text{for $E < E_b$} \\[0.3 cm]
 K_0 \;  2^{ \Delta \alpha \, (w_R-w_L)} \; 
 \left ( \frac{E}{E_0} \right )^{-\alpha_1} \;
 \left [1 + \left (\frac{E}{E_b} \right )^{\frac{1}{w_R}} \right ]^{-\Delta\alpha \, w_R} &
 \text{for $E > E_b$} ~,
\end{cases}
 \label{eq:break-parw2} 
\end{equation}
so that the spectral index $\alpha(E)$ takes the form:
\begin{equation}
 \alpha(E) = \begin{cases} 
 \overline{\alpha} + \frac{\Delta \alpha}{2} \;
 \tanh \left [ \frac{\ln (E/E_b)}{2 w_L} \right ] & 
 \text{for $E < E_b$} \\[0.25 cm]
 \overline{\alpha} + \frac{\Delta \alpha}{2} \;
 \tanh \left [ \frac{\ln (E/E_b)}{2 w_R} \right ] & 
 \text{for $E > E_b$~}  ~.
\end{cases}
 \label{eq:alpha-parw2} 
\end{equation}
For this parametrization the flux and its first derivative
(i.e. the spectral index) are continuous, but the second derivative is
discontinuous at the point $E = E_b$.
Taking the derivative of the spectral index with respect to energy,
one finds that the limits for $E \to E_b$ taken in the two directions are different:
\begin{equation}
\lim_{E \to (E_b)^\mp} \frac{d \alpha(E)}{d \ln E} = \frac{\Delta \alpha} {4 \, w_{L,R}} ~.
\end{equation}
This appear to be a tolerable  flaw for the parametrization of Eq.~(\ref{eq:break-parw2}).

\begin{figure}[bt]
\begin{center}
\includegraphics[width=11.0cm]{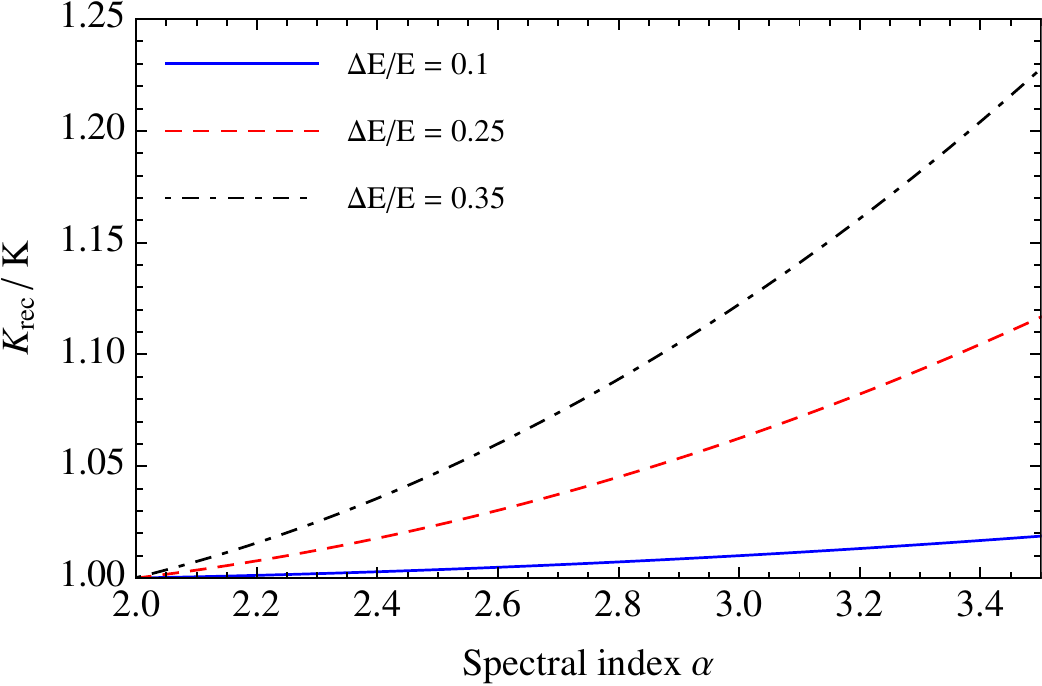}
\end{center}
\caption {\footnotesize
  Effects of the detector   finite resolution  (assumed  gaussian with constant $\Delta E/E$)
  on the normalization  of a power law  spectrum.
  The curves show the ratio $K_{\rm rec}/K$  plotted as  a function of the spectral index $\alpha$ for
  three values  of the resolution:  $\Delta E/E = 0.1$, 0.25, 0.35. 
\label{fig:plopow} }
\end{figure}

Having constructed   this more  general  parametrization of a spectral feature,
we have tested  that  for the level of  precision of the existing data,
the   form   that depends on a single  width is  in fact  adequate  to describe all
known  structures  (see discussion in the following).

\section{Detector resolution}
\label{sec:resolution}
The shape of an observed spectral feature is distorted by the detector
resolution. To illustrate these instrumental effects one can consider
a simple example where the energy of the events is reconstructed with
gaussian errors and a constant $\Delta E/E$.
With this assumption a spectrum that is an unbroken power law
$\phi(E) = K \, E^{-\alpha}$ results (in the absence of an unfolding)
in a reconstructed spectrum that is a power law with the same exponent:
$\phi_{\rm rec} (E_{\rm rec}) = K_{\rm rec} \; (E_{\rm rec})^{-\alpha}$. The only effect
is a modification of the constant $K_{\rm rec}$, with a ratio
$K_{\rm rec}/K$ that is a function of the spectral index $\alpha$ and the
resolution $f = \Delta E/E$:
\begin{equation}
 \frac{K_{\rm rec}}{K} = g(\alpha,f) =
 \frac{1}{\sqrt{2 \pi} \, f} \; \int_0^\infty dx~x^{\alpha-1} ~\exp \left [-\frac{1}{2 \, f^2} (1 -x)^2
 \right ]~.
\end{equation}
A  graphics representation of this  function is shown in Fig.~\ref{fig:plopow}.
For a spectral index $\alpha > 2$ the factor $g(\alpha, f)$ is larger than unity,
reflecting the fact that the ratio
$\langle E\rangle/E_{\rm rec} < 1$, where 
$\langle E\rangle$ is the average true energy
of events of reconstructed energy $E_{\rm rec}$.
This is a simple consequence of the fact that 
the spectrum is rapidly falling with energy.
The effect becomes more important when the resolution is poor
(growing with $f$) and when the spectrum is steep (growing with $\alpha$)
but remains always rather small.
For example $g(2.7, 0.2) \simeq 1.024$, $g(2.7,0.3) \simeq 1.054$
$g(3.0, 0.3) \simeq 1.090$.

\begin{figure}[bt]
\begin{center}
\includegraphics[width=14.0cm]{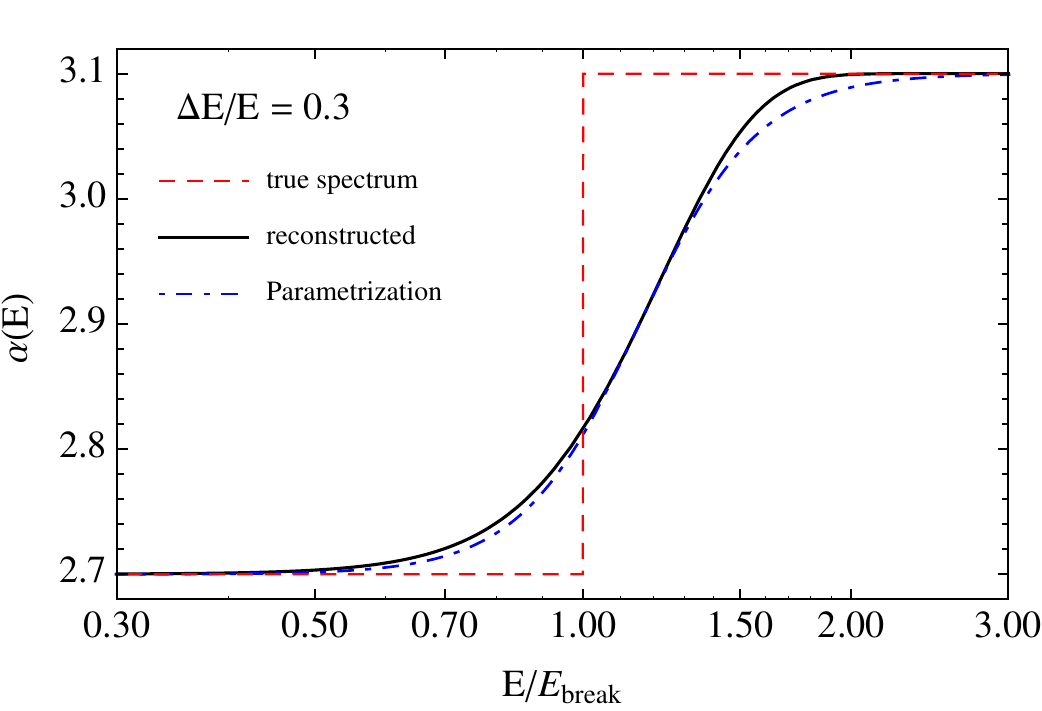}
\end{center}
\caption {\footnotesize
Effects of detector resolution on the shape of the reconstructed energy spectrum.
In this example the true spectrum has a sharp knee-like frature
at energy $E_{\rm break}$ where the spectral index (indicated by the red, dashed line)
changes abruptly from 2.7 to 3.1. 
The solid line shows the spectral index of the measured spectrum calculated
assuming Gaussian errors in the reconstruction of
the energy of the events with a resolution $\Delta E/E = 0.3$.
The (blue) dot--dashed line is the parametrization of the spectral index
of Eq.~(\ref{eq:index-parametrization}) for the ``optimum choice''of parameters
(see main text).
 \label{fig:plorec0}}
\end{figure}

A spectral break with vanishingly small width ($w = 0$) at energy $E_{\rm b}$ will
be observed as a feature with a finite width and a shape that reflects the
detector resolution. An example of the spectral index
of the experimentally reconstructed flux (without unfolding) is shown
in Fig.~\ref{fig:plorec0}.

\begin{figure}[bt]
\begin{center}
\includegraphics[width=10.cm]{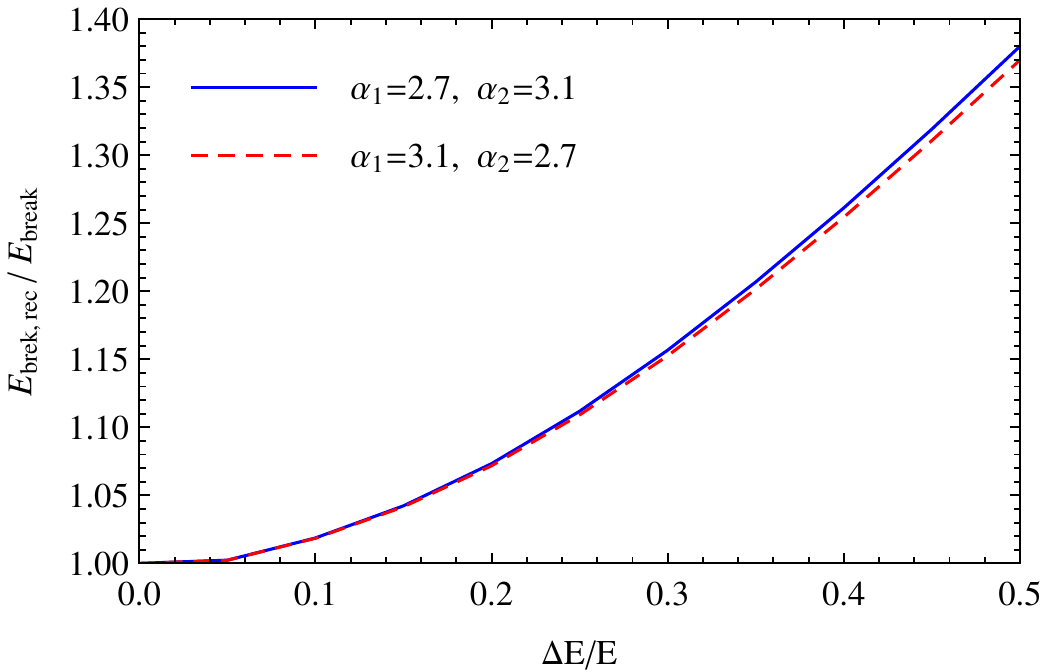}
\end{center}
\caption {\footnotesize
Effects of detector resolution on the shape of a sharp spectral feature
The errors on the energy reconstruction are assumed Gaussian with a constant $\Delta E/E$.
The curves show the ratio between
reconstructed and the true break energy,  plotted as as a function of the resolution
$\Delta E/E$.  The solid curve is for a softening feature
(with $\alpha_{1} = 2.7$ and $\alpha_{2} = 3.1$). The dashed curve  for a
hardening feature  (with $\alpha_{1} = 3.1$ and $\alpha_{2} = 2.7$). 
\label{fig:plorec1} }
\end{figure}

The observed shape is similar, but not
exactly equal to the form of the parametrization in Eq.~(\ref{eq:index-parametrization}).
The asymptotic values of the spectral index for $E\to 0$ and $E \to \infty$ are equal
to the true ones, but the energy 
where the slope of the reconstructed flux is equal to the average value
$(\alpha_2+\alpha_1)/2$ is not $E_b$ but has a value $E_{b,{\rm rec}}> E_b$.
This can easily understood as a consequence of the fact already discussed that
$\langle E \rangle/ E_{\rm rec} < 1$.
An example of this shift in the position of the break energy, 
Fig.~\ref{fig:plorec1}   shows the ratio
$E_{b,{\rm rec}}/E_b$ plotted as as a function of the detector resolution $\Delta E/E$.
For a discontinuity of order $\Delta \alpha \simeq 0.4$ the 
shift is a factor of order 1.07 for a resolution of 20\% and 1.15 for a resolution
of 30\%.

\begin{figure}[bt]
\begin{center}
\includegraphics[width=11.0cm]{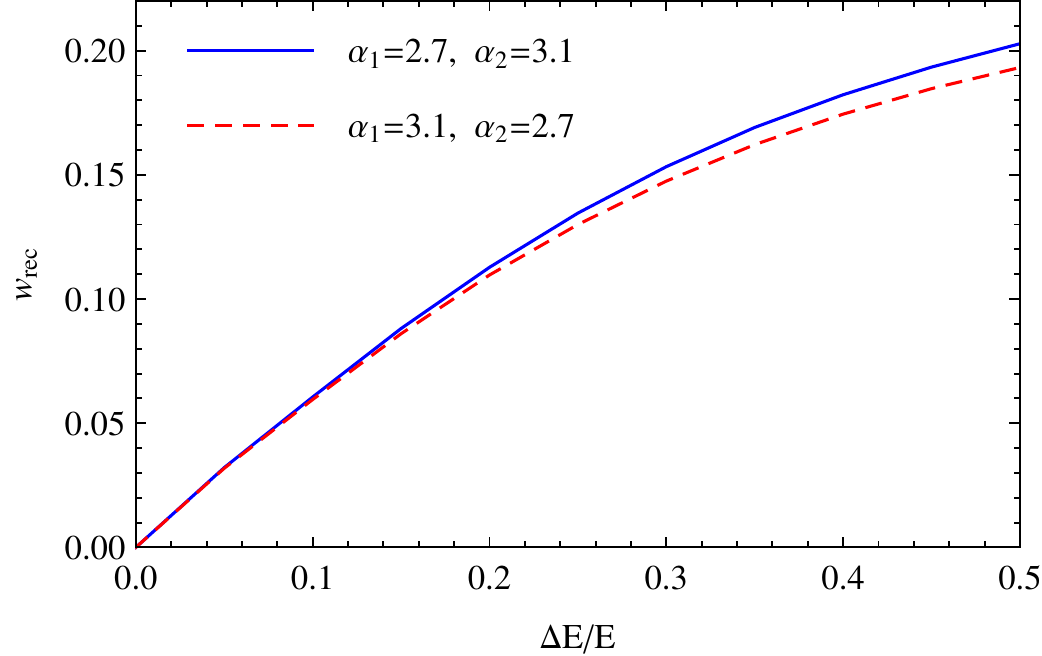}
\end{center}
\caption {\footnotesize
As in Fig.~\ref{fig:plorec1}.
The curves show the reconstructed width $w_{\rm rec}$ as a function of the
detector resolution.
The solid (dashed) line corresponds to a softening (hardening) spectral feature.
\label{fig:plorec2} }
\end{figure}

The detector resolution has also the effect that a very narrow spectral
feature is reconstructed as a more gradual softening (or hardening).
This effect is illustrated in 
Fig.~\ref{fig:plorec2} that shows the reconstructed width $w_{\rm rec}$
as a function of the detector resolution.
The quantity $w_{\rm rec}$ is estimated from Eq.~(\ref{eq:weff}) as the width
of the energy range where one half of the jump  $\Delta \alpha$  develops.
For a  sharp ($w =0$) break  with $\Delta \alpha \simeq 0.4$
one finds that $w_{\rm rec}$ is of order 0.11 for a resolution of 20\% and of order 0.15
for a resolution of 30\%.
The important point here is that, in the
absence of systematic effects, is it very unlikely to
observe spectral features that are narrower than the width generated by the
detector resolution.

For completeness it should be also added that the detailed shape
of an observed spectral feature
generated by the detector resolution effects acting on a very narrow
structure is not identical
(see Fig.~\ref{fig:plorec1})  to the parametrization of Eq.~(\ref{eq:break-parametrization})
as it is not exactly symmetric in $\log E_{\rm rec}$, since the evolution in the range
$E_{\rm rec} < E_{b, {\rm rec}}$ is slightly faster
than the evolution in the range $E_{\rm rec} > E_{b, {\rm rec}}$ (see Fig.~\ref{fig:plorec1}).
This effect is however small and in most cases it can be safely neglected. 

In the more general (and realistic case) where
a spectral feature has a true width, it is obviously necessary to
convolute the detector effects with the real shape of the structure.

\section{Two component flux}
\label{sec:2-components} 
A simple and natural interpretation of a hardening feature in the cosmic ray spectrum
is that the flux is formed by the sum of two components that are reasonably
well described in the energy region of the break by power law spectra.
In this  scenario the  flux has the form:
\begin{equation}
 \phi(E) =
 K_1 \; \left ( \frac{E}{E_0} \right )^{-\alpha_1} 
+ K_2 \; \left ( \frac{E}{E_0} \right )^{-\alpha_2} 
\end{equation}
(with $E_0$ again an arbitrary reference energy).  The two components  will be equal
at one  (unique) crossing energy:
\begin{equation}
E_{\rm cross} = E_0 \; \left (\frac{K_1}{K_2} \right )^{1/(\alpha_1 - \alpha_2)}
\end{equation}
The total flux can be then rewritten in the form:
\begin{equation}
 \phi(E) = K_1 \; \left ( \frac{E}{E_0} \right )^{-\alpha_1} \;
 \left [1 + \left ( \frac{E}{E_{\rm cross}} \right )^{-(\alpha_2 - \alpha_1)} \right ] ~
\end{equation}
This form corresponds exactly to the parametrization of Eq.~(\ref{eq:break-parametrization})
for the  value of the width:
\begin{equation}
w = \frac{1}{\alpha_1 - \alpha_2}~
\label{eq:2-components}
\end{equation}
(without loss of generality one can assume that $\alpha_1 > \alpha_2$, so
that the first component is the softest one, and therefore
$w = (\alpha_1 - \alpha_2)^{-1} > 0$).
Eq.~(\ref{eq:2-components}) states the (very intuitive) result
that the width of a spectral feature that corresponds
to the transition between  components that have power law form depends on the
difference between the spectral indices of the two components, and becomes broader 
when the two exponents are close to each other.
This  result can be used to test the hypothesis   that a hardening feature
is the manifestation of the  transition  between two  components
that are unbroken    power laws
(see the discussion in the following).

\section{Spectral features in the CR spectrum}
\label{sec:features}
In this section we will very briefly discuss the  shape of some of the most
prominent features in the flux of  protons and   of the all--particle spectrum.

We will consider here only the  energy range $E \gtrsim 30$~GeV.
The discussion of  cosmic rays at low energy
where the spectra exhibit large and energy dependent curvature,
and are also distorted by  time dependent solar modulation effect  is
an important topic,  but it will  covered  here.
We also  will not discuss the suppression of the CR flux at the
highest energies ($E \sim 10^{20}$~eV).

\subsection{The Cream/Pamela ``discrepant hardening''}
An intriguing hardening feature is present in the spectra
of  protons and helium  (and other nuclei) at a rigidity of order 300~GV.
The first indication of this hardening emerged indirectly, 
from a comparison of the spectra measured by
the CREAM balloon experiment \cite{Ahn:2010gv}
in the energy range 1--10$^3$~TeV, with the spectra measured at lower
energy by magnetic spectrometers such as Caprice \cite{Boezio:2002ha},
BESS \cite{Haino:2004nq} and AMS01 \cite{Aguilar:2002ad}.
The CREAM collaboration  noted   that to connect
their measurements of the proton and helium spectra to the
lower energy  data it was  necessary to  assume  the existence of 
a ``discrepant hardening''  in the spectra. 

\begin{table}
  \caption{\footnotesize Parameters of fits to the proton spectrum  in  the energy range 
    around  1~TeV. The last line is the fit  to the AMS02 and CREAM  data performed
    in this  work. The fits are shown in Fig.~\ref{fig:ploprotfita}.
    \label{tab:protons}}
\begin{center}
  \renewcommand{\arraystretch}{1.5}
  \begin{tabular}{ | l || c |  c |  c | c |}
\hline
Experiment   &  $R_b$  (GeV) &  $\alpha_1 $   &  $\alpha_2$  & $w$   \\
\hline
PAMELA \cite{Adriani:2011cu}
&  $232^{+35}_{-30}$  &   $2.85\pm0.04$  &  $2.67 \pm 0.05$ &  $\approx 0$    \\
AMS02 \cite{Aguilar:2015ooa}
&  $336^{+95}_{-52}$  
&  $2.849^{+0.005}_{-0.006}$  &  $2.716^{+0.075}_{-0.062}$  &  $0.18^{+0.27}_{-0.18}$  \\   
CREAM \cite{Yoon:2017qjx}     &  --   & -- &  $2.61 \pm 0.01$  &   -- \\
\hline
Combined fit & $1010 \pm 70$  & $2.794 \pm 0.003$  & $2.57 \pm 0.04$ & $0.41 \pm 0.04$ \\
\hline
  \end{tabular}
\end{center}
\end{table}

This prediction  received an important  confirmation
from the measurements of  the proton and helium  spectra  performed
by PAMELA \cite{Adriani:2011cu} that observed  
hardenings in both  spectra at a  rigidity
of order 230--240~GV.

\begin{figure}[bt]
\begin{center}
\includegraphics[width=14.0cm]{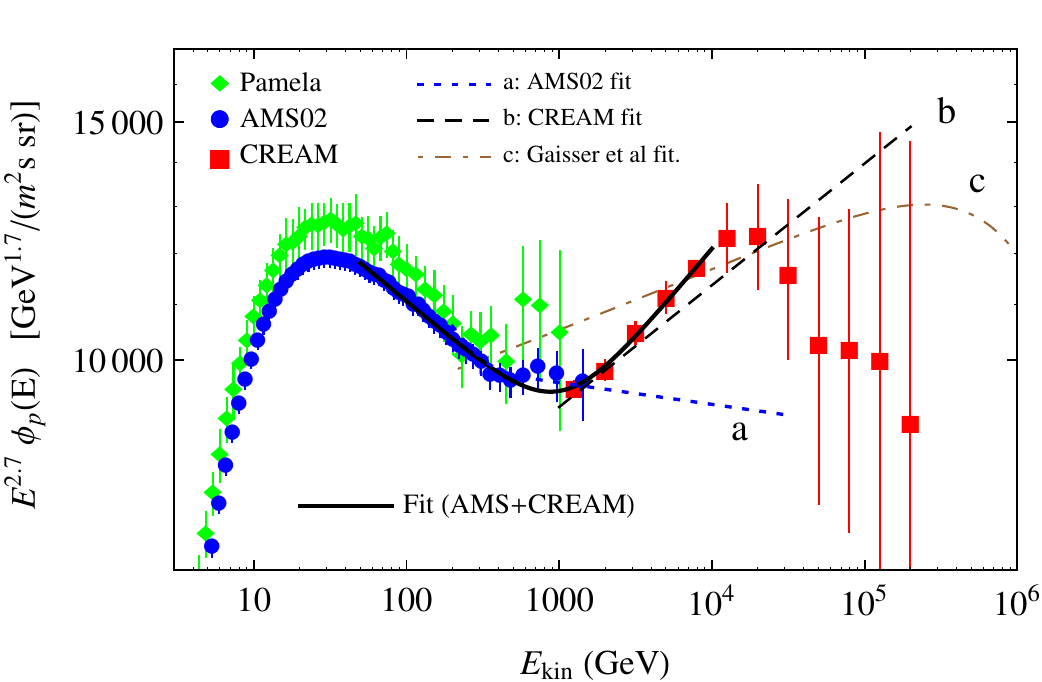}
\end{center}
\caption {\footnotesize
Measurements of the proton flux of Pamela (diamonds) \protect\cite{Adriani:2011cu})
AMS02 (circles) \protect\cite{Aguilar:2015ooa} and CREAM (squares)
\protect\cite{Yoon:2017qjx}.
The dashed line is the fit of the AMS02 data in the original publication.
The solid line is a fit of the combined AMS02 and CREAM data (see main text).
The dot--dashed line is the proton flux
estimated by Gaisser et al. \protect\cite{Gaisser:2013bla}
\label{fig:ploprotfita}}
\end{figure}

Later, the  AMS02 detector \cite{Aguilar:2015ooa,Aguilar:2015ctt}
has also measured the proton and helium spectra with higher precision
and in a sligthly broader rigidity range, 
%that extends to 1.8~TV for protons and 3.0~TV for helium.
confirming  the existence of the hardenings of the two spectra
but measuring a spectral shape not identical to what was  obtained by PAMELA.
The hardenings measured by AMS02 are 
centered at higher  rigidity,  have a smaller  $\Delta \alpha$,
and a broader width  (see table~\ref{tab:protons}).

The   data  of CREAM, PAMELA and AMS02 on the proton spectrum
are shown in Fig.~\ref{fig:ploprotfita}.
In the case of CREAM   the points  refer to a recent publication 
\cite{Yoon:2017qjx} that includes the  observations of a second long duration  ballon flight.
The three  collaborations have performed  fits to their data
that are listed  in table~\ref{tab:protons}.   The  PAMELA   collaboration  \cite{Adriani:2011cu}
fits the data with  is  broken  power law  form (that is a   feature with width $w=0$).
The AMS02 fit uses the parametrization of  Eq.~(\ref{eq:break-par-ams}),
and in table~\ref{tab:protons}  the parameter  $w$ is  estimated using Eq.~(\ref{eq:s-w}).
The CREAM  collaboration \cite{Yoon:2017qjx} 
fits their  data with a  simple  power law, obtaining a spectral  index  $\alpha_p \simeq 2.61 \pm 0.01$.
Inspecting Fig.~\ref{fig:ploprotfita}   one can however note
that there are some indications  of a softening of the spectrum for  $E \gtrsim 10$~TeV,
so that  a  fit to the data  limited  to  the 1--10~TeV   energy range
would yield a smaller value of the spectral index.

Comparing the  data of the three experiments 
one can  notice that the 
spectral  index measured by CREAM ($\alpha_p \simeq 2.61$)
is  significantly smaller that the asymptotic (high  energy) spectral  indices
fitted by PAMELA  ($\alpha_p \simeq  2.67$) and AMS02 ($ \alpha_p \simeq 2.71$).
This  suggests  the  possibility that the    hardening feature in the proton spectrum
is very broad, and extends  beyond the rigidity range of the two magnetic  spectrometers,

To explore this  possibility, we  have performed a fit of the AMS02 and CREAM data
in the energy range from 50 to $10^4$~GeV (a total of 34 data points)
using the 5 parameter form of Eq.~(\ref{eq:break-parametrization})
that describe  a  single spectral  feature.
The combination of the AMS02 and CREAM data  can be well  described by the parametrization
of Eq.~(\ref{eq:break-parametrization}).   Combining quadratically  statistical and  stystematic
errors,  one obtains  $\chi^2_{\rm min}= 4.9$
(this very small value suggests the existence
of significant correlations between the systematic errors  for data points at different energies).
The best fit parameters are 
$E_b = 1011\pm 70$~GeV, $\alpha_1 \simeq 2.79 \pm 0.03$, $\alpha_2 \simeq 2.57 \pm 0.04$ and
$w \simeq 0.41 \pm 0.04$ (the errors have been estimated using  $\chi^2 < \chi^2_{\rm min} + 5$).
This exercise suggests that it is likely that the proton hardening around one TeV is in fact
a very broad feature that extends from 200~GeV to 2 TeV.

The correct description of the proton flux in the energy range 10--100~TeV,
is also of great importance  as  a boundary condition for the studies of the CR flux in
the Knee region. Fig.~\ref{fig:ploprotfita} also shows  the fit  to the proton flux
performed by Gaisser et al. \cite{Gaisser:2013bla}   taking into account the
measurements of the extensive air shower  detectors at higher energy.

A discussion of the  flux of helium and other nuclei  in this energy range
is of course very important,  but  it is postponed to a future  work.

\subsection{The ``Knee''}
The  prominent structure of the  ``Knee'' in the all--particle spectrum
at an energy of order 3~PeV  has attracted  much attention.
It is   obvious  that to obtain a full  understanding of the  origin of the knee it
is essential to measure  separately the energy distributions of  the different components
(protons, helium nuclei, $\ldots$)  that form  the  spectrum.
Estimates of the spectra of  different components   have been  in fact obtained 
for example by Kascade \cite{Antoni:2005wq}, ARGO--YBJ \cite{Bartoli:2015vca},
and Kascade--GRANDE \cite{Apel:2013uni,Apel:2013ura,Apel:2011mi},  however
the  determination of the primary particle mass in air shower detectors is  difficult
and the systematic uncertainties (mostly associated to the modeling of hadronic interactions)
are large and poorly understood.  A problem  of  great importance is that
the estimates of the proton spectrum
in the PeV region  by the Kascade and ARGO--YBJ  detectors  are  not in  good agreement.

\begin{figure}[bt]
\begin{center}
\includegraphics[width=14.0cm]{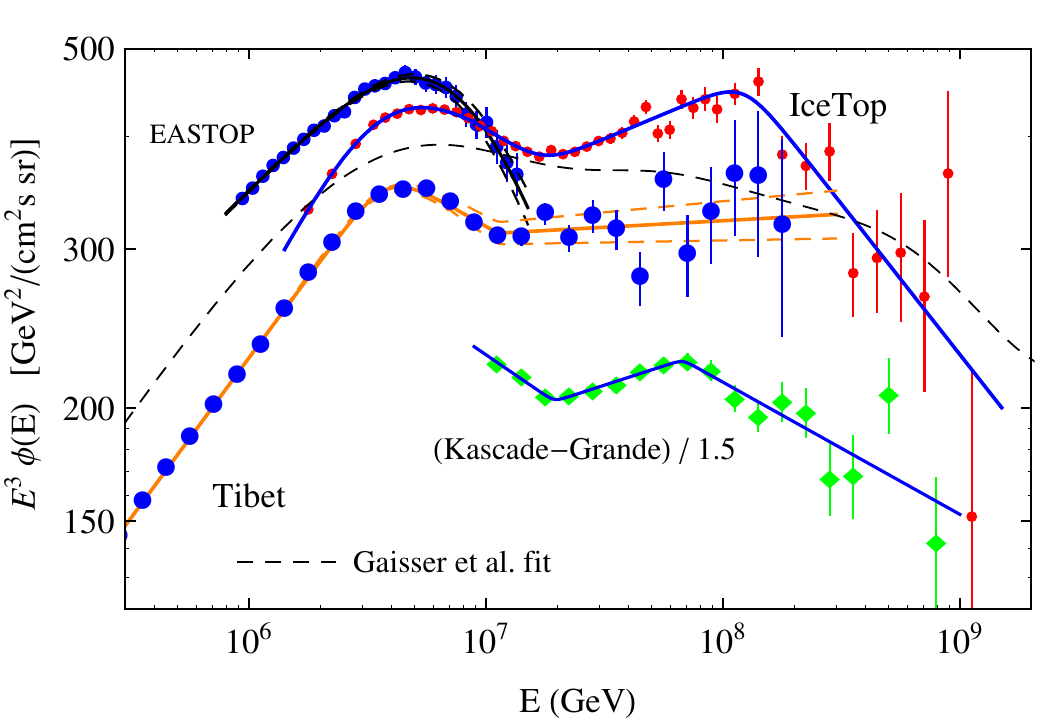}
\end{center}
\caption {\footnotesize
Measurements of the all particle CR spectrum
in the energy region $E = 10^6$--10$^9$~GeV by
EAS--TOP \cite{Aglietta:1998te},
TIBET \cite{Amenomori:2008aa}, 
Kascade-Grande \cite{Apel:2012tda} and IceTop \cite{Aartsen:2013wda}.
The lines are fits to the individual measurements
using the parametrization of a spectral feature of Eq.~(\ref{eq:break-parametrization}).
\label{fig:ploknee}}
\end{figure}

For these reasons it  remains interesting to study the detailed  shape of the  all--particle
spectrum.  Figure~\ref{fig:ploknee} shows some selected measurements of the all--particle spectrum in the
energy region from 1 to $10^3$~PeV.
The data  shown is from  EAS--TOP \cite{Aglietta:1998te},
TIBET \cite{Amenomori:2008aa}, Kascade-Grande \cite{Apel:2012tda} and IceTop \cite{Aartsen:2013wda}.
The Tibet experiment has presented in \cite{Amenomori:2008aa} 
three different estimates of the CR flux, obtained using different assumptions
for the hadronic interaction model and the particle composition, and the 
the spectrum shown in Fig.~\ref{fig:ploknee} is the one estimated
using the Sibyll interaction model \cite{Ahn:2009wx}).

Inspection of Fig.~\ref{fig:ploknee}  shows the presence of  important differences
between the measurements of the different experiments that are the
manifestation of  the existence of large  systematic uncertainties.
In fact, a  detailed study of  how the  differences in the reconstructed  spectra
are related  to different methods of measurement, 
different  models of shower  development,
and  different assumptions on the chemical composition of the CR flux,
could  yield  very important information.

Even  in the presence of these systematic effects,
the measurements of the all--particle flux in the
1--30~PeV  energy range   reveal the existence of some  interesting structure
in the  shape of the spectrum,  that
appears  to  have not one, but  two  features: 
a gradual  softening centered  at $E \simeq 3$--4~PeV,
followed by a smaller width hardening  at $E \simeq 10$--15~PeV.

The spectra of the different experiments can be  well  fitted   assuming the existence
of these two  features.
A list of the best fit parameters is given in table~\ref{tab:knee}.
Note that the  EASTOP detector \cite{Aglietta:1998te}
covers only the lower  energy part  of the knee  region, and only  observes
the spectral softening, while  the Kascade--Grande  detector \cite{Apel:2012tda}
covers only the higher energy region, and observes only the hardening.
The fits are also shown in Fig.~\ref{fig:ploknee}.

It is possible that these two (softening and  hardening)  features   in the all--particle spectrum
have a distinct origin, however  given how  close they are,
it seems  more likely that a  physical model  that  explain
these structures  will  have to address them together,
and that what is  commonly called ``the Knee'' should be considered as
the combination of these two substructures.

In  Fig.~\ref{fig:ploknee}  we also show the model of the all--particle spectrum constructed
by Gaisser et al.  \cite{Gaisser:2013bla}, where the  spectrum of Galactic cosmic rays is
modeled  as the combinations of three populations of sources
that release  power law spectra of particles  that  have rigidity dependent exponential
cutoffs at energy  3, 30 and 2000 ($\times Z$)~PeV.  This type of  models  is  adequate
to  describe the main features of the spectrum, but  the two
Knee substructures are not accurately reproduced. 

\begin{table}
  \caption{\footnotesize Fits to the all--particle CR spectra in the energy range 1--30~PeV.
    The spectrum is  described as  three
    segments   where spectral index is approximately constant, with values
    $\alpha_1$, $\alpha_2$ and $\alpha_3$, separated
    by two spectral  features   with break energy $E_{b1}$ and $E_{b2}$ and widths
    $w_1$ and $w_2$.
    \label{tab:knee}}

  \vspace{0.2 cm}
{\footnotesize  (a) Parameters for the lower energy (softening) feature.  }
  
\begin{center}
\begin{tabular}{ | l || c |  c |  c | c |}
\hline
Experiment   &  $E_{b1}$  (PeV)  &  $\alpha_1 $   &  $\alpha_2$  &   $w_1$  \\
\hline
EAS--TOP   &  $7.7 \pm 0.3$  &  $2.73 \pm 0.01$  &  $3.75 \pm 0.06$  &  $0.44 \pm 0.02$  \\
TIBET      &  $3.7 \pm 0.6$ &  $2.62 \pm 0.01$   &  $3.16 \pm 0.02$  &  $0.18 \pm 0.02$  \\   
IceTop     &  $3.7 \pm 0.3$ &  $2.50 \pm 0.01$   &  $3.18 \pm 0.01$  &  $0.35 \pm 0.01$  \\
\hline
\end{tabular}
\end{center}

  \vspace{0.2 cm}
{\footnotesize  (b) Parameters for the higher energy (hardening) feature. }

  \begin{center}
\begin{tabular}{ | l || c |  c |  c | c |}                                                
\hline
Experiment   &  $E_{b2}$  (PeV)  &  $\alpha_2 $   &  $\alpha_3$  &   $w_2$   \\   
\hline
TIBET             &  $11.2 \pm 0.2$ &   $3.16 \pm 0.02$  &   $2.98 \pm 0.07$ &  $< 0.15$  \\
IceTop            &  $17.5 \pm 0.7$ &   $3.18 \pm 0.01$  &   $2.89 \pm 0.02$ & $0.13 \pm 0.01$ \\
Kascade--Grande   &  $19.4 \pm 0.7$  &   $3.18 \pm 0.01$  &  $2.91 \pm 0.02$ &  $< 0.15$ \\
\hline
\end{tabular}
\end{center}
\end{table}

\subsection{From the ``Knee''to the ``Ankle''} 

The all--particle spectrum  in the energy range  between the Knee and the Ankle
cannot be well fitted     as a simple unbroken   power law, and there  is  
evidence  that the spectrum  undergoes some  softening.

A structure  that  has  received  a lot of attention is the so called ``second Knee'',
a softening  feature  observed   at $E \approx  300$~PeV
by several  experiments. This  spectral feature  is potentially of great significance
because it has  been  identified as possibly  marking the
transition between  Galactic  and extragalactic cosmic rays \cite{Aloisio:2006wv}.

A review of Bergman and Belz \cite{Bergman:2007kn}
summarizes  early  results  of Akeno, Fly's Eye and HiRes, obtaining a global best fit
for a sharp  break  centered at $E \simeq  330\pm 15$~PeV     where the spectral index
undergoes  a  change $\Delta \alpha \simeq 0.25 \pm 0.013$.

More recent data however are not  entirely  consistent with these conclusions.
The Telescope Array  Low energy extension (TALE)  has recently   reported   \cite{Ivanov:2015pqx}
a  softening at an energy just below  previous estimates  $E \simeq 200 \pm 23$~PeV, but softening
features at lower energy are  present in the data of 
Kascade-Grande \cite{Apel:2012tda}  (centered at $E \simeq 67\pm 8$~PeV)
and IceTop  \cite{Aartsen:2013wda} (centered at $E \simeq 123 \pm 15$~PeV)
The Kascade--Grande and IceTop  data, together with our fits  are  shown in Fig.~\ref{fig:ploknee}.
The HiRes \cite{Abbasi:2005ni} and
Telescope Array  \cite{TheTelescopeArray:2015mgw}  data, 
together with  our  fits  centered at  480 and 300~PeV are
shown in Fig.~\ref{fig:ploankle},

It should be added   that  if  the transition between Galactic and extragalactic  cosmic rays
is indeed  in the region between the Knee and the Ankle, it is virtually certain
that it must have a  shape   that is  not  well fitted by a simple  formula
such as Eq.~(\ref{eq:break-parametrization}).

In fact, a priori one expects that the  Galactic/extragalactic  transition   should
correspond  to a  spectral hardening, simply because 
the extragalactic flux,   that  emerges  as  dominant  above the transition energy $E^*$,
out of hypothesis,  must be  harder  that the Galactic  flux,
however there is no  significant spectral hardening  in the energy range 20--4000~PeV.

The transition  can correspond to a softening  only if  three
special  conditions  are satisfied. \\
(i) The Galactic component  has a softening feature at  $E_{\rm break}^{\rm Gal} \approx E^*$. \\
(ii) Also the extragalactic component undergoes a  softening
(presumably with a different astrophysical origin) for  $E_{\rm break}^{\rm extra} \approx E^*$. \\
(iii) The  two components are normalized so that 
$\phi_{\rm Gal} (E^*) \approx \phi_{\rm extra} (E^*)$. \\
If these three conditions are satisfied it is then possible to obtain that
for $E \lesssim E^*$ the CR flux is dominated by the Galactic component
before it udergoes its softening; 
for $E \gtrsim E^*$ the flux is dominated by the extragalactic component,
after its  softening;  and around the transition energy $E \approx E^*$  
one  observes  a reasonably smooth  spectral  softening. 

As an example, one can  consider a simple  model where  the  CR  Galactic component
is a power law of exponent $\alpha_0$  with an  exponential
(or quasi-exponential) cutoff at  $E^*_{\rm Gal}$:
\begin{equation}
 \phi_{\rm Gal} (E) \simeq K_{\rm Gal} ~\left ( \frac{E}{E_0} \right )^{-\alpha_0}
 ~\exp \left [-\frac{E}{E^*_{\rm Gal}} \right ] 
\end{equation}
while the extragalactic flux has a ``knee--like'' feature at energy $E^*_{\rm extra}$:
\begin{equation}
 \phi_{\rm extra} (E) \simeq K_{\rm extra} ~\left ( \frac{E}{E_0} \right )^{-\alpha_1}
 ~ \left [1 + \left (\frac{E}{E^*_{\rm extra}} \right )^{\frac{1}{w}} \right ]^{-(\alpha_2 - \alpha_1)\, w}
\end{equation}
If $E^*_{\rm Gal} \approx E^*_{\rm extra} \simeq E^*$ and
$K_{\rm extra}/K_{\rm  Gal} \approx  (E^*/E_0)^{\alpha_1 -\alpha_0}$
(so that the two components are approximately equal  for $E \simeq E^*$)
and if the spectral indices  are ordered as:
$\alpha_1 < \alpha_0, < \alpha_2$,  the total flux  will appear in first approximation
as   having a softening  at $E \simeq  E^*$  where  the spectral index   changes from
$\alpha_0$ (the exponent of the  Galactic  component before its  cutoff) to
$\alpha_2 > \alpha_0$ (the exponent  of the extragalactic  component  at high energy).
A scenario similar to the one outlined above is in fact a the basis of the
``Dip model'' of Berezinsky and collaborators \cite{Aloisio:2006wv}. 

A study  of such  scenario however shows that
a sufficienly accurate measurement  of the shape  of the spectrum around the transition
should show  significant deviations  from a simple  form such as the parametrization
of Eq.~(\ref{eq:break-parametrization}).

\subsection{The ``Ankle''} 
The existence of an  hardening of the all--particle  spectrum at $E \approx  4$--5~EeV  has been
established already in  the 1990's  by Fly's Eye and Akeno, and these results  has been then  
later  confirmed  by Haverah  Park,  Yakutksk,  AGASA, HiRes and more recently by Auger and Telescope Array
(for  reviews  see \cite{Bergman:2007kn,Bluemer:2009zf,Matthiae:2015dma}.

The interpretation of this  spectral  feature has already generated a  large body of  literature.
A possibility is that  it  marks the Galactic/extragalactic  CR transition, an alternative
\cite{Aloisio:2006wv} is that it is a ``dip''   created by energy losses effects on
a flux of extragalactic protons.

\begin{figure}[bt]
\begin{center}
\includegraphics[width=14.0cm]{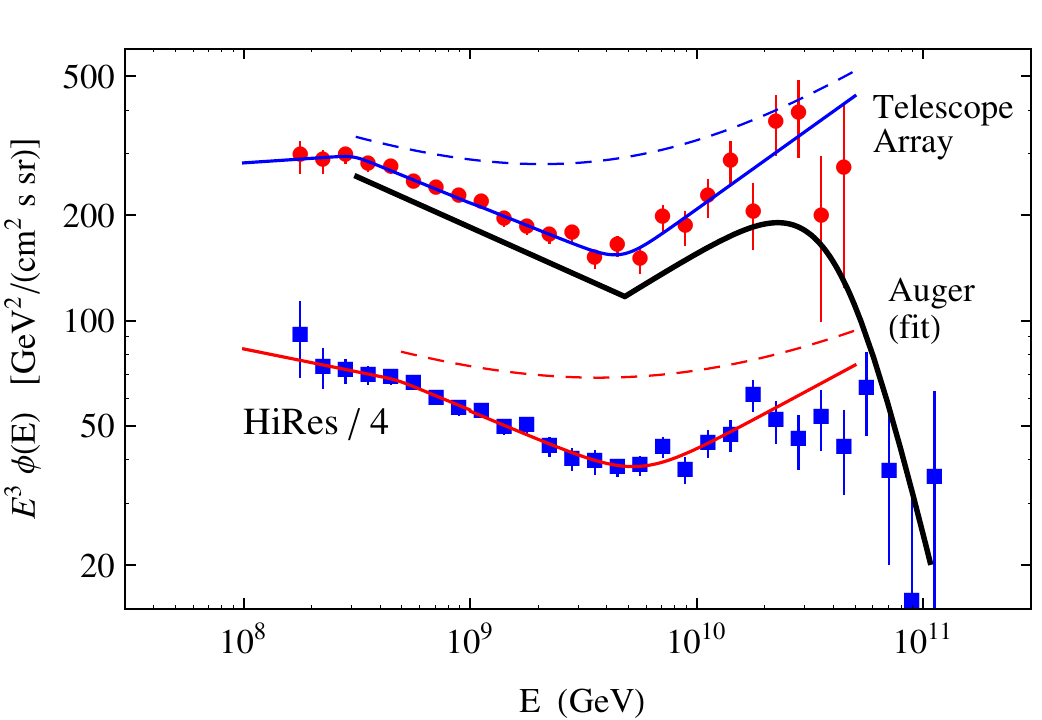}
\end{center}
\caption {\footnotesize
Measurements of the all particle CR spectrum
in the energy region $E = 10^8$--10$^{11}$~GeV by
 Telescope Array (circles) \cite{TheTelescopeArray:2015mgw}
and HiRes (squares, divided by a factor 4) \cite{Abbasi:2005ni}.
The thick line is the fit of the Auger data \cite{Aab:2015bza}.
The dashed  lines  show the flux  obtained summing the two
asymptotic (for low and high energy)  power law spectra
obtained  with the best fit [$\phi(E) = K_1 \; E^{-\alpha_1} +K_2 \; E^{-\alpha_2}$].
\label{fig:ploankle}}
\end{figure}

Fig.~\ref{fig:ploankle} shows the fit  of the spectral  shape of the all--particle
spectrum  performed   by Auger  \cite{Aab:2015bza}, and the data  of the
HiRes \cite{Abbasi:2005ni}
and  Telescope Array \cite{TheTelescopeArray:2015mgw}
(note that the HiRes data is  rescaled by a factor 1/4).
The spectra  measured  by the three experiments 
exhibit an evident  hardening  at $E \approx  4$--5~EeV.
This   spectral  feature  can be  well  described  by the  parametrization
of Eq.~(\ref{eq:break-parametrization}).
In fit of the Auger collaboration (see table~\ref{tab:ankle})  describes
the Ankle as  a zero--width   spectral break.
The results of our fits to the HiRes and Telescope array,
performed in the energy range  0.5--40~EeV,    are also listed 
in table~\ref{tab:ankle})  and shown   in Fig.~\ref{fig:ploankle}  and yield  widths of order 0.23 and 0.12.

\begin{table}
  \caption{\footnotesize Parameters of fits to the all--particle spectrum in the
    Ankle region.  The fits are shown in Fig.~\ref{fig:ploankle}.
    \label{tab:ankle}}
\begin{center}
  \renewcommand{\arraystretch}{1.5}
{  %\footnotesize
  \begin{tabular}{ | l || c |  c |  c | c |}
\hline
Experiment   &  $E_{a}$  (EeV)  &  $\alpha_1 $   &  $\alpha_2$  & $w$   \\
\hline
Auger \cite{Aab:2015bza}
&  $4.82 \pm 0.07 \pm 0.8$  &  $3.29 \pm 0.02 \pm 0.05$  &  $2.60\pm 0.02 \pm 0.1$   &  $\approx 0$ \\   
HiRes
&  $5.5^{+0.6}_{-0.4}$  &  $3.27 \pm 0.01$  &  $2.65\pm 0.07$  &  $0.23 \pm 0.04$  \\   
Tel. Array
&  $4.7^{+0.5}_{-0.4}$  &  $3.25 \pm 0.01$  &  $2.53\pm 0.09$ &  $0.12 \pm 0.02$   \\   
\hline
  \end{tabular}
}
\end{center}
\end{table}

The main point we  would like to stress  here is that the Ankle
is observed as a  narrow spectral feature by all experiments,
(with stimate of the width of  order 0.1--0.3).
This result is incompatible with the simplest hypothesis that
the spectrum is the superposition of two power law components.
As  discussed in section~\ref{sec:2-components},
in this  case the width should be $w \simeq |\Delta \alpha| \approx 1.3$--1.6,
that is  one order of magnitude  larger.
As an illustation in Fig.~\ref{fig:ploankle} the two dashed lines show
the flux obtained as the sum $\phi(E) = K_1 \; E^{-\alpha_1} + K_2 \; E^{-\alpha_2}$
where the two components
($K_1 \; E^{-\alpha_1}$ and $K_2 \; E^{-\alpha_2}$) are the asymptotic forms
of the fitted spectrum for energy much smaller and much larger than
the Ankle energy. The flux obtained in this way is larger, and 
has a spectral index that changes much more slowly than the data.

This result does not exclude the hypothesis that the Ankle
is created by the transition between regimes where the CR flux
is dominated by two different components (for example Galactic and extragalactic
particles), however it excludes the possibility that the two components
have a simple, unbroken power law form.

\section{Conclusions}
The identification, and  detailed experimental  study of  the spectral  features
in the energy distributions  of cosmic rays is an essential  tool to
develop  an  understanding of the astrophysical  mechanisms of
acceleration and propagation  that  determine the fluxes.

A  measurement of the width of a  spectral  feature, that is the range of
$\log E$   where the  change in spectral index  develops,
can be a very  important constraint  for  models  that  want to
interpret  the CR spectra. It  can be useful  to have  a  standard 
definition   for a parameter  that measures  this 
property of the spectral features  that is  commonly accepted.
In this work we  have suggested the use of a simple and very natural   definition
for the  width of  a  spectral feature: $w \simeq \Delta \log_{10} E$  (or more precisely
$w = \Delta \log_{10} E/\log_{10} 9$)  where
$\Delta \log E$  is the  range of $\log E$   where one half of the
jump in spectral index of the feature develops.

As an illustration  we have discussed three examples of  spectral  features.
The first example is the hardening   of the proton flux at $E \simeq 300$~GeV.
We find that a  simultaneous   fit of the AMS02 and CREAM data suggests 
that this  spectral structure is very broad ($w \simeq 0.4$) and 
centered at high energy ($E_b \simeq 1$~TeV).
The  second example we considered is the well known  Knee in the all--particle spectrum
at $E \approx 3$--4~PeV. We suggest  that this
spectral feature should be seen as  formed  by two
substructures, an extended softening
at $E \simeq 4$~PeV with a width  $w \simeq 0.2$--0.4,  followed  at
$E \simeq 15$--20~PeV by a hardening of narrower width.
The third example is the  Ankle at $E \simeq 4$~EeV  that has a   quite small width
$w \simeq 0.1$--0.25.  This  implies  that if the  Ankle marks the transition
between Galactic and extragalactic  cosmic rays, it is  necessary to  assume
that at leat one of the two components has  significant structure  around
the transition energy, because the combination of two  unbroken power laws
would result in a much broader  spectral feature.

If the Galactic to extragalactic transition is  below the Ankle and corresponds
to a softening of the all particle spectrum one expects to find a shape
of the energy distribution  that, if measured  with  sufficient accuracy, is  not
well described by a simple  form, because   both (Galactic and extragalactic)
components must have  a non trivial  energy dependence.

\end{document}